\documentclass[useAMS,usenatbib]{mn2e}

\usepackage{times}
\usepackage{color}

\title[Power requirements for cosmic ray re-acceleration]{Power requirements for cosmic ray propagation models involving re-acceleration
and a comment on second order Fermi acceleration theory}
\author[A. Thornbury and L. O'C. Drury]{Andrew Thornbury$^{1}$
and Luke O'C. Drury$^{2}$\thanks{E-mail ld@cp.dias.ie http://orcid.org/0000-0002-9257-2270}\\
$^{1}$School of Physics, Trinity College, The University of Dublin, Dublin 2, Ireland\\
$^{2}$Dublin Institute for Advanced Studies, 31 Fitzwilliam Place, Dublin 2, Ireland}
\begin{document}

\date{}

\pagerange{\pageref{firstpage}--\pageref{lastpage}} \pubyear{2002}

\maketitle

\label{firstpage}

\begin{abstract}
We derive an analytic expression for the power transferred from interstellar turbulence to the Galactic cosmic rays in propagation models which include re-acceleration.  This is used to estimate the power required in such models and the relative importance of the primary acceleration as against re-acceleration.  The analysis provides a formal mathematical justification for Fermi's heuristic account of second order acceleration in his classic 1949 paper.
\end{abstract}


\section{Introduction}

The importance of re-acceleration for cosmic ray transport in the Galaxy has been debated inconclusively for many years \citep[among many others]{1986ApJ...300...32S,1988ICRC....7...87C,1994ApJ...431..705S, 2010ApJ...725.2110S,2011ApJ...729..106T}
but is widely thought to be important and is routinely implemented, for example, in the de facto standard numerical model GALPROP \citep{2013AAS...22240102M}.  At some level it is clear that re-acceleration, which is just the second-order Fermi acceleration associated with the same interstellar turbulence that produces the spatial diffusion, must occur.  The question is to what extent this modifies the observed spectra and how much energy is involved. Remarkably this last aspect, the power required to drive the re-acceleration or equivalently, its contribution to the total cosmic ray luminosity of the Galaxy, appears to have attracted little attention.  This has motivated us to write this short article.

\section{Fundamentals of Galactic cosmic ray diffusive transport}

At least since the seminal monograph of \cite{1969ocr..book.....G} the standard model for cosmic ray transport in the Galaxy has been based on the idea that the cosmic ray particles are scattered by irregularities in the interstellar magnetic field and execute, on sufficiently large scales, a form of random walk or diffusion process until they eventually escape from the Galaxy.  This diffusion occurs both in physical space and in momentum space so that the full transport equation contains two diffusion terms (along with many other terms representing other physical processes such as ionisation energy losses, nuclear interactions, advection etc.)
\begin{equation}
{\partial f\over\partial t} = Q + 
\nabla\left(D_{xx}\nabla f\right) +  {1\over 4\pi p^2} {\partial\over\partial p}\left(4\pi p^2 D_{pp} {\partial f\over\partial p}\right) +....
\end{equation}
Here $f(x,p,t)$ is the isotropic part of the phase space density of a given species as a function of position $x$, scalar momentum $p$ and time $t$ and $Q(x, p, t)$ is a source term representing the initial production and acceleration of cosmic rays, generally thought to occur in supernova remnants, {e.g.\ } \citet{2012APh....39...52D}.  The spatial and momentum space diffusion coefficients are $D_{xx}$ and $D_{pp}$ respectively and will in general be functions of position, momentum, and time. 

If the same physical scattering process produces both diffusion terms, it is clear that the two diffusion coefficients must be related.  The relationship is best understood using a simple heuristic model where we assume that particles of a given energy or momentum are isotropised on a scattering time scale $\tau$.  If the particles have velocity $v$ this means that the mean free path is of order $\lambda = \tau v$ and thus, from kinetic theory,
\begin{equation}
D_{xx} \approx {1\over 3} {\lambda^2\over\tau} = {1\over 3} \lambda v = {1\over 3} {v^2 \tau}.
\end{equation}
If the scattering is off structures in the magnetic field moving at a velocity of order the Alfv\'en speed $V_A$ (these could be actual Alfv\'en waves, but it is probably better to just think of $V_A$ as the characteristic velocity associated with the propagation of disturbances in the field) then on each scattering the particle will suffer a random change in momentum of order
\begin{equation}
\Delta p \approx p {V_A\over v}
\end{equation}
and thus
\begin{equation}
D_{pp} \approx {1\over 3} {(\Delta p)^2\over \tau}
= {1\over 3} p^2 {V_A^2\over v^2\tau}.
\end{equation}
It follows that 
\begin{equation}
D_{xx}D_{pp} \approx {1\over 9} p^2 V_A^2
\end{equation}
and the two coefficients are inversely related to one another which is physically understandable; if there is very strong scattering, the particles diffuse little in space, but a lot in momentum and vice versa.

Much more elaborate calculations yield essentially the same result \citep{2002cra..book.....S}, for example \cite{1975MNRAS.172..557S} shows that for two populations of forward and backward propagating Alfv\'en waves with scattering rates $\nu_+$ and $\nu_-$ 
\begin{equation}
D_{xx} = v^2 \left<1-\mu^2\over 2(\nu_+ + \nu_-)\right>
\end{equation}
and 
\begin{equation}
D_{pp} = 4\gamma^2 m^2 V_A^2 \left< {1-\mu^2\over 2} {\nu_+\nu_-\over \nu_+ + \nu_-}\right>
\end{equation}
where the angle brackets denote an averaging over the pitch angle distribution $\mu$ being the pitch angle cosine.  Noting that in Skilling's notation
\begin{equation}
p^2 = \gamma^2 m^2 v^2
\end{equation}
it follows that, if we assume equal rates of forward and backward scattering with $\nu_+=\nu_- = \nu$,
\begin{equation}
D_{xx}D_{pp} =
p^2 V_A^2 \left<1-\mu^2\over2\right>^2 = {1\over 9} p^2 V_A^2
\end{equation}
for isotropic pitch angle distributions.  Of course if the rates of forward and backward scattering are not quite equal, or the distributions not quite isotropic, the precise numerical factor will vary somewhat.  The key physical point however is that the two coefficients are inversely related and that their product is of order $p^2 V_A^2/9$.

\section{The re-acceleration power}

How much energy is transferred per unit time and volume from the interstellar turbulence to the cosmic rays through this process?  If we write $T(p)$ for the kinetic energy corresponding to particles of (scalar) momentum $p$, then the energy density of the cosmic rays is just
\begin{equation}
E_{CR}=\int_0^\infty 4\pi p^2 f(p) T(p) dp
\end{equation}
and from the basic diffusive transport equation,
\begin{equation}
{\partial E_{CR}\over\partial t} = ... + \int_0^\infty 4\pi T{\partial\over\partial p}\left(4\pi p^2 D_{pp} {\partial f\over\partial p}\right) \,dp...
\end{equation}
The re-acceleration power, the quantity we wish to evaluate, is thus
\begin{equation}
P_R = \int_0^\infty 4\pi T{\partial\over\partial p}\left(4\pi p^2 D_{pp} {\partial f\over\partial p}\right) \,dp.
\end{equation}

For reasonably behaved distribution functions which have a high-energy cut-off, and diffusion coefficients and distributions which do not have pathological behaviour at very low energies, we can integrate twice by parts to get
\begin{eqnarray}
P_R &=& \left[4\pi T p^2 D_{pp} {\partial f\over\partial p}\right]_0^\infty - 4\pi \int_0^\infty p^2 v D_{pp} {\partial f\over \partial p} \,dp\\
&=& - 4\pi \int_0^\infty p^2 v D_{pp} {\partial f\over \partial p} \,dp\\
&=& \left[4\pi p^2 v D_{pp} f\right]_0^\infty + \int_0^\infty f {\partial\over\partial p}\left(4\pi p^2 D_{pp} v\right)\, dp\\
&=& \int_0^\infty 4\pi p^2 f {1\over p^2} {\partial\over\partial p} \left(p^2 D_{pp} v\right) \, dp
\end{eqnarray}
were we have used the relativistic identity $dT = v dp$.

We now substitute for $D_{pp}$ in terms of $D_{xx}$ and rearrange terms to get the interesting result
\begin{eqnarray}
P_R &=& \int_0^\infty 4\pi p^2 f {1\over p^2} {\partial\over\partial p} \left(p^4V_A^2 v\over 9 D_{xx}\right) \, dp\\
&=& \int_0^\infty 
4\pi p^2 f \left (V_A^2 p v\over 9 D_{xx}\right) \left[ 4 + {\partial\ln(v/D_{xx})\over\partial\ln p}\right].
\end{eqnarray}

In this expression $pv = T$ for relativistic particles and $2T$ for non-relativistic particles and the logarithmic slope term is small for any reasonably power-law dependence of $D_{xx}$ on $p$,
\begin{equation}
\left|{\partial\ln(v/D_{xx})\over\partial\ln p}\right| <1.
\end{equation}
In fact with the conventional parametrisation $D_{xx}\propto v p^\delta$ the final bracket is just $[4-\delta]$ and in the generally favoured reacceleration models $\delta = 0.3$.

Thus to order of magnitude the re-acceleration time scale, 
\begin{equation}
{E_{CR}\over P_R} \approx {9\over 4} {D_{xx}\over V_A^2}
\end{equation}
as was to be expected on dimensional grounds.

\section{Relevance to Fermi's 1949 theory}

Another interesting way to write the above expression is in terms of the scattering time $\tau$.  Substituting $D_{xx}=v^2\tau/3$ we get
\begin{equation}
P_R = \int_0^\infty 
4\pi p^2 f \left (V_A^2 p v\over 3 v^2 \tau\right) \left[ 4 + {\partial\ln(v/D_{xx})\over\partial\ln p}\right]
\end{equation}
which can be interpreted as saying that the average energy gain per scattering of each particle is just
\begin{equation}
\Delta T = pv {4\over 3} {V_A^2\over v^2} \left[ 1 + {1\over 4}{\partial\ln(v/D_{xx})\over\partial\ln p}\right].
\label{deltaT}
\end{equation}
Ignoring the small correction for the power-law dependence of the scattering represented by the last bracket this is basically the same as the result originally proposed on heuristic grounds in the seminal paper of \cite{1949PhRv...75.1169F}, that there is an energy gain on each scattering which is the particle's energy times a factor of order the square of the ratio of the scattering centre velocity to the particle velocity.  In this sense the result can be seen as validating Fermi's physical intuition and showing how a pure diffusion term in the transport equation can be physically equivalent to an energy gain per scattering which is quadratic in the velocity ratio.  The result is even more transparent in terms of the mean change in scalar momentum, $\Delta p = \Delta T/v$,
\begin{equation}
{\Delta p\over p} = {4\over 3} {V_A^2\over v^2} \left[ 1 + {1\over 4}{\partial\ln(v/D_{xx})\over\partial\ln p}\right].
\end{equation}

At first sight this is a rather remarkable result, because the diffusion term can transfer energy in either direction.  It is therefore surprising that it should be formally equivalent to a fixed energy gain per scattering.  The key of course is that the derivation relies on the distribution function having regularity properties which allow the twofold integration by parts to be made without contributions from the end-points.  This excludes pathological distribution functions which grow without limit at high energies.  It is interesting that even if the diffusion flux is reversed at some momenta (flowing from high energies to low energies), on average this must be balanced out by the contributions from other parts of the distribution function.

It is also interesting to note that the result relies only on there being a pure diffusion process in momentum space.  There is no need to invoke a higher rate of "head-on" collisions as against "tail-on" collisions, as in Fermi's paper.  There will be a second-order Fermi acceleration even if the rates are exactly equal simply because there is more phase-space volume at high energies.  To put it another way, a random walk, on average, moves you away from the origin even if there is no directional bias to the individual steps.    Fermi's physical intuition led him to the right answer, but the explanation he gave was incorrect.

\section{Numerical estimates of time scales and power}

If we use the `standard' inferred value for the average spatial diffusion in the Galaxy \citep{2011ApJ...729..106T} of $D_{xx}\approx 7\times 10^{28}\,\rm cm^2s^{-1}$ at a particle rigidity of $4\,\rm GV$ and an Alfv\'en speed of order $V_A\approx 40\,\rm km\, s^{-1}$ the re-acceleration time-scale comes out as being $3\times 10^8\,\rm yr$ so that, for conventional confinement times of particles at these energies of $3\times 10^7\,\rm yr$ we are looking at a 10\% effect.  This is consistent with the idea that the re-acceleration modifies the secondary to primary ratios at these energies, but is not so large that it swamps the main effect which is energy-dependent escape.  However in view of the uncertainties in the parameters the effect, especially at lower energies, could be much larger and contribute a significant amount of the cosmic ray luminosity of the Galaxy with clear implications also for the energy budget of the interstellar turbulence.   

Unfortunately we have little knowledge of the interstellar cosmic ray proton spectrum an energies below $1\,\rm GeV$, but it is clear from the form of the re-acceleration power integral that care is needed if one is not to require unreasonable amounts of power at these mildly sub-relativistic energies.  The problem is clear from equation (\ref{deltaT}) which shows that the mean energy transfer grows rapidly as the particles become sub-relativistic (as $v^{-2}$).  Thus if the re-acceleration is to have a 10\% effect at energies of a few GeV, it is likely to be totally dominant at sub-GeV energies and may in fact represent the major energy input into this part of the spectrum.  It would be desirable to study this further with numerical simulations such as GALPROP as well as with constraints on the low-energy cosmic ray flux from interstellar chemistry and the ionisation of molecular clouds.  While re-acceleration models reduce the energy required for the primary injection of cosmic rays, this may be more than out-weighed by the re-acceleration power requirements.

\section*{Acknowledgments}
This paper derives from a final year project undertaken as part of the Theoretical Physics course in TCD by the first author under the supervision of the second.

\label{lastpage}


\begin{thebibliography}{99}

\bibitem[\protect\citeauthoryear{Cesarsky}{1988}]{1988ICRC....7...87C} 
Cesarsky C., 1988, ICRC, 7, 87 

\bibitem[\protect\citeauthoryear{Drury}{2012}]{2012APh....39...52D} Drury 
L.~O'C.~., 2012, APh, 39, 52 

\bibitem[\protect\citeauthoryear{Fermi}{1949}]{1949PhRv...75.1169F} Fermi 
E., 1949, PhRv, 75, 1169 

\bibitem[\protect\citeauthoryear{Ginzburg 
\& Syrovatskii}{1963}]{1969ocr..book.....G} Ginzburg V.~L., Syrovatskii S.~I., 1963, The Origin of Cosmic Rays, Izd. Akad. Nauk. SSSR, Moscow.  Authorised English translation, Pergamon Press, Oxford, 1964.

\bibitem[\protect\citeauthoryear{Moskalenko 
\& GALPROP Team}{2013}]{2013AAS...22240102M} Moskalenko I., GALPROP Team, 2013, AAS, 222, \#401.02 

\bibitem[\protect\citeauthoryear{Schlickeiser}{2002}]{2002cra..book.....S} 
Schlickeiser R., 2002, Cosmic Ray Astrophysics, Springer Verlag, Berlin. 

\bibitem[\protect\citeauthoryear{Seo 
\& Ptuskin}{1994}]{1994ApJ...431..705S} Seo E.~S., Ptuskin V.~S., 1994, ApJ, 431, 705 

\bibitem[\protect\citeauthoryear{Shalchi \& B{\"u}sching}{2010}]{2010ApJ...725.2110S} Shalchi A., B{\"u}sching I., 2010, ApJ, 725, 2110 

\bibitem[\protect\citeauthoryear{Simon, Heinrich, 
\& Mathis}{1986}]{1986ApJ...300...32S} Simon M., Heinrich W., Mathis K.~D., 1986, ApJ, 300, 32 

\bibitem[\protect\citeauthoryear{Skilling}{1975}]{1975MNRAS.172..557S} 
Skilling J., 1975, MNRAS, 172, 557 

\bibitem[\protect\citeauthoryear{Trotta et al.}{2011}]{2011ApJ...729..106T} 
Trotta R., J{\'o}hannesson G., Moskalenko I.~V., Porter T.~A., Ruiz de 
Austri R., Strong A.~W., 2011, ApJ, 729, 106 


\end{thebibliography}
\end{document}